# Asymmetric Price Adjustment over the Business Cycle

**Daniel Levy** [a,b,c,d,e] *****, **Haipeng (Allan) Chen** [f], **Sourav Ray** [g], **Elliot Charette** [h,i], **Xiao Ling** [j], **Weihong Zhao** [k], **Mark Bergen** [l], **Avichai Snir** [a]

[a] *Department of Economics, Bar-Ilan University, Ramat-Gan 5290002, Israel*
[b] *Department of Economics, Emory University, Atlanta, GA 30322, USA*
[c] *International Centre for Economic Analysis, Wilfrid Laurier University, Waterloo, Ontario, Canada*
[d] *International School of Economics at Tbilisi State University, 0108 Tbilisi, Georgia*
[e] *The Rimini Centre for Economic Analysis (RCEA)*
[f] *Tippie College of Business, University of Iowa, Iowa City, IA 52242-1994, USA*
[g] *G.S. Lang School of Business and Economics, University of Guelph, Guelph, ON N1G 2W1, Canada*
[h] *Department of Applied Economics, University of Minnesota, Saint Paul, MN 55108, USA*
[i] *Federal Reserve Bank of Minneapolis, Minneapolis, MN 55401, USA*
[j] *School of Business, Central Connecticut State University, New Britain, CT 06050, USA*
[k] *Smith School of Business, University of Maryland, College Park, MD 20742, USA*
[l] *Carlson School of Management, University of Minnesota, Minneapolis, MN 55455, USA*

**Last Revision**: June 11, 2025

### Abstract

Studies of micro-level price datasets find more frequent *small* price increases than decreases, which can be explained by consumer inattention because time-constrained shoppers might ignore small price changes. Recent empirical studies of the link between shopping behavior and price attention over the business cycle find that consumers are more (less) attentive to prices during economic downturns (booms). These two sets of findings have a testable implication: the asymmetry in small price changes should vary over the business cycle—it should diminish during recessions and strengthen during expansions. We test this prediction using a large US store-level dataset with more than 98 million weekly price observations for the years 1989–1997, which includes an 8-month recession period, as defined by the NBER. We compare price adjustments between periods of recession (high unemployment) and expansion (low unemployment). Focusing on small price changes, we find, consistent with our hypothesis, that there is a greater asymmetry in small price changes during periods of low unemployment compared to the periods of high unemployment, implying that firms' price-setting behavior varies over the business cycle.

# 1. Introduction

Many studies that focus on the effects of aggregate conditions on price-setting decisions focus on asymmetric price adjustment.[1] However, studies of the effects of aggregate conditions on the asymmetry in *small* price changes are scarce. To fill this gap in the literature, we study how asymmetry in small price changes varies over the business cycle.

There is evidence of asymmetry in small price changes: there are more frequent small price increases than decreases. The evidence comes from Spain, France, Israel, the USA, Norway, Brazil, the European Union, etc., for prices of food, computers, camera equipment, etc.[2] For example, Chen et al. (2008) study weekly prices of a large US food chain with more than 98 million observations for 1989–1997, and find more frequent small price increases than decreases for price changes of up to 10¢–15¢, about 5% of the average price. See Figure 1 (which is similar to Figure 2 in Chen et al., 2008).[3]

As an explanation, Chen et al. (2008) offer consumer inattention. Time-constrained consumers who buy dozens of goods might be inattentive to small price changes because paying attention to current prices and comparing them to last period's prices is time-consuming and cognitively demanding (Shugan 1980, Reis 2006a and 2006b, Mankiw and Reis 2010, Sayag et al. 2025). For example, according to Sims (1998, pp. 320-321), "Because individuals have many things to think about and limited time, they can devote only limited intellectual resources to the tasks of data-gathering and analysis. We know from personal experience that many data that we could look up daily, and that are in principle relevant to our optimal economic decision-making, do not in fact influence our behavior, *except when they change dramatically*, or perhaps when we occasionally set aside some time to re-assess our portfolio" (our emphasis). Similarly, Reis (2006b, p. 1761) observes that "[Consumers] …face costs of acquiring, absorbing and processing information… [They] rationally choose to only sporadically update their information and recompute their optimal…plans. In between updating dates, they remain inattentive." Therefore, the cost of processing information on small price changes might exceed the benefit, creating a range of inattention along the demand curve, where the consumer is inattentive to small price changes.

---

[1] See, for example, Blinder (1991), Hannan and Berger (1991), Mankiw and Romer (1991), Lach and Tsiddon (1992, 1996, 2007), Ball and Mankiw (1994), Blinder et al (1998), Estrada and Hernando (1999), Fisher and Konieczny (2000, 2006), Peltzman (2000), Mankiw and Reis (2002), Álvarez and Hernando (2004), Davis and Hamilton (2004), and Rotemberg (2005).

[2] See Buckle and Carlson (2000), Álvarez and Hernando (2004), Baudry et al. (2004), Cecchetti (2004), Rátfai (2004, 2006, 2007), Ellingsen, et al. (2005), Ray et al. (2006, 2012), Konieczny and Rumler (2007), Lach and Tsiddon (2007), Vermeulen et al. (2007), Chen et al. (2008), Klenow and Kryvtsov (2008), Barros et al. (2009), Konieczny and Skrzypacz (2010), Reis (2010), Klenow and Malin (2011), Midrigan (2011), Wulfsberg (2016), and Sayag et al. (2024). Eichenbaum et al. (2014), Cavallo and Rigobon (2016), and Cavallo (2018) argue that small price changes could be due to measurement errors. Even these studies, however, find a non-negligible share of small price changes that cannot be explained by measurement errors.

[3] Chen et al. (2008) report that their finding is robust. For example, the asymmetry is also present in low-inflation and deflation periods, it holds for alternative measures of inflation and for products whose prices have not increased, and it is robust to lagged price adjustment.

This range of inattention makes small price *decreases* less valuable to the retailer because slightly lower prices don't trigger a response from consumers who do not notice the small price cut. A small price *increase*, however, is valuable exactly for the same reason—consumers won't notice a small price rise and therefore will not reduce purchases. Thus, the retailer has an incentive to make more frequent small price increases than decreases.[4]

It follows that asymmetry should vary with consumer attention: If consumers have more (less) time, and thus are more (less) attentive, we'll see less (more) asymmetry. Business cycles offer an opportunity to test this prediction, because studies of shopping behavior find a correlation between unemployment and attention: times of high (low) unemployment coincide with greater (less) attention to prices. Thus, high (low) unemployment should coincide with lower (greater) asymmetry in small price changes. That is, as unemployment varies over the cycle, we should see a corresponding variation in consumer attention, and thus, in asymmetry in small price adjustments.

We test this prediction using Chen et al.'s (2008) data, exploiting the fact that their sample period, 1989–1997, contains an 8-month NBER recession. We compare asymmetry in small price changes between the highest and lowest unemployment periods and find that it is indeed stronger when the unemployment rate is low, suggesting that firms' price-setting behavior varies over the business cycle.

Next, we discuss the link between shopping time and price attention. Section 3 discusses testable predictions. Section 4 describes the data. Section 5 presents the findings. Section 6 assesses the robustness. Section 7 concludes by summarizing the findings, noting some caveats, and suggesting directions for future research.

## 2. Shopping time and price attention over the business cycle

The observation that opportunity costs affect price search is well-established. According to Becker (1965, p. 516), "…the unemployed…would be more willing to spend their time in a queue…than would high-earning males." In Stigler's (1961) model, low-income families can cut their expenditures by greater price-attentiveness and search.

Studies of shopping-time variation over the business cycle are consistent with these predictions. Long et al. (2015) find that households pay lower prices when unemployment is high, and that in the Great Recession, when unemployment rose from 5.8% to 9.6%, households adjusted shopping intensity and became more price attentive.

---

[4] As Chen et al. (2008, p. 735) note, "The possibility that consumers may be inattentive to price changes is consistent with the observation that retailers alert the public about promotions by posting sale signs, to ensure shoppers *notice* the price discounts" (emphasis in original).

Nevo and Wong (2015) report that in the Great Recession, as unemployment rose, families looked for more deals, made more frequent shopping trips, shopped more at discount stores, used more coupons, and bought more private-label products. Indeed, the share of private label products is counter-cyclical.[5] Cha et al. (2015) report that in the Great Recession, Americans cut restaurant visits and bought more grocery items. In addition, they find that higher unemployment led households to purchase cheaper brands at cheaper outlets.

Aguiar and Hurst (2007) find that older households (whose opportunity cost of time is lower) pay lower prices, shop more intensively, and use more coupons than middle-aged households. Aguiar et al. (2013) find increased shopping-related activities in the Great Recession. McKenzie et al. (2011) find that in the post-2002 crisis, Argentinians increased shopping time and frequency, concluding that "…increase in shopping search is one of the most prevalent adjustment mechanisms used by Argentinian consumers to cope with the crisis" (p. 3).

Lastly, recent studies of informational rigidities also find that people are more attentive in recessions than in expansions (Coibion and Gorodnichenko 2015, Goldstein 2023).

## 3. Consumer inattention and cyclical variation in price adjustment asymmetry

The consumer inattention argument implies the following hypothesis: during high (low) unemployment, the opportunity cost of time is lower (higher), and thus people are more (less) attentive to small price changes. Also, the value of being attentive to small price changes increases (decreases) in periods of high (low) unemployment. Thus, higher (lower) unemployment would coincide with greater (reduced) attention and therefore, with a lower (higher) asymmetry in small price changes.

We follow Chen et al. (2008, p. 730) to define an asymmetry threshold as "…the last point at which the frequency of price increases exceeds the frequency of price decreases of the same absolute magnitude $(z \geq 1.96)$." In the absence of asymmetry, there should be an equal number of price increases and decreases for each size of price change. Consumer inattention to small price changes implies that we should see more small price increases than small price decreases. The variation in consumer inattention over the business cycle (e.g., unemployment) predicts smaller asymmetry thresholds in periods of high unemployment compared to low unemployment periods.

## 4. Data

We use the same data as Chen et al. (2008), scanner data from Dominick's, a large chain that was operating in Chicago with 94 stores. The data contains 400 weekly prices over 8 years, Sept. 14, 1989–

---

[5] See Volpe (2011, 2014), Dube et al (2018), Marmorstein et al (1992), Hoch and Banerji (1993), Quelch and Harding (1996), Lamey et al (2012).

May 8, 1997, a total of 98,691,750 observations for 18,037 products, in 29 categories.[6]

The 8-year sample period contains an 8-month NBER recession period, from August 1990 to March 1991, which we exploit for comparing the asymmetry during the recession and expansion.

A key macro determinant of consumer attention, as noted above, is unemployment. However, because unemployment lags output by two quarters, the highest and the lowest unemployment periods do not coincide exactly with the recession and expansion periods, respectively.

Therefore, we conduct two analyses. First, we compare the asymmetry thresholds for the *NBER recession months* (capturing the high unemployment effect) with the asymmetry thresholds for the lowest unemployment months. Second, we compare the asymmetry thresholds for the *highest unemployment months* with the asymmetry thresholds for the lowest unemployment months.

We repeat the analysis twice. First, we use the average US unemployment rate to determine the highest and the lowest unemployment rate periods. Second, we use the Chicago unemployment rate to determine the highest and the lowest unemployment rate periods. The latter is useful as Dominick's was operating in the Chicago area. We run the analyses using 8-month windows because the NBER recession in our sample was 8 months long.

As Figure 2 indicates, the period of the lowest US unemployment rate coincides with that of Chicago and occurs in September 1996–April 1997, averaging 4.8% and 5.2%, respectively. The highest unemployment rate period, according to the *u*-US series, is from February 1992–September 1992, averaging 7.6%, while according to the *u*-Chicago series, it is from December 1991–July 1992, averaging 8.1%. During the NBER recession months, the unemployment rate averaged 6.3%.

## 5. Empirical findings

In the LHS panel of Table 1, we report the asymmetry thresholds.[7] We exclude from the analyses two product categories, beers and cigarettes, because the products included in these categories are highly regulated (Chen et al., 2008, footnote 2). The cross-category average asymmetry threshold for the lowest unemployment period, $\bar{A} = 10.30¢$, is larger than for the NBER recession period $\bar{A} = 0.62¢$, for the Highest-*u* Chicago period $\bar{A} = 4.15¢$, and for the Highest-*u* US period, $\bar{A} = 3.59¢$.

Across the 27 categories, in 62 out of 75 comparisons (82.7% of cases), we find a stronger asymmetry for the lowest unemployment period, in 5 cases (6.7% of cases) we find equal asymmetry, and in 8 cases (10.7% of cases), we find weaker asymmetry for the lowest unemployment period than for the other periods (Chakraborty et al. 2015). The theoretical prediction of inattention is statistically supported: 82.7% > 50% with $z = 5.65$, $p < 0.0001$. A paired $t$-test confirms this conclusion: for the 27 product categories, the asymmetry is larger for the lowest unemployment period ($\bar{A} = 10.30$, SD = 7.99) than for the other three periods ($\bar{A}_{NBER} = 0.62$, SD = 1.02, $t_{20} = 5.18$, $p < 0.001$; $\bar{A}_{Chicago} = 4.15$, SD = 4.75, $t_{26} = 3.43$, $p < 0.005$; $\bar{A}_{us} = 3.59$, SD = 4.09, $t_{26} = 3.91$, $p < 0.001$).

---

[6] For more details about the data, see Dutta et al. (2002), Barsky et al. (2003), Chen et al. (2008), Levy et al. (2002, 2010, 2011, 2020), Snir et al. (2022), and Dominick's Data User Manual, available at https://www.chicagobooth.edu/-/media/enterprise/centers/kilts/datasets/dominicks-dataset/dominicks-manual-and-codebook_kiltscenter, accessed March 23, 2025. The data is publicly available and it can be downloaded from the homepage of the University of Chicago's Booth School of Business, www.chicagobooth.edu/research/kilts/datasets/dominicks, accessed March 23, 2025.

[7] There is only one "Lowest-*u*" column because, as noted above, the periods in which the US and Chicago unemployment rates attain the lowest average values over an 8-month period coincide.

We find 2.5–16.6 times stronger asymmetry when unemployment is low. Thus, the data is consistent with the consumer inattention hypothesis, irrespective of the criterion used for identifying the high-unemployment period.

## 6. Robustness

The finding is unlikely to be driven by sample size differences. Although the lowest unemployment period has a larger sample size than the other three periods, the statistical significance of the differences is not high. A paired $t$-test of the sample size averages yields the following results: lowest-$u$ vs NBER recession, $t_{20} = 2.52$, $p < 0.02$; lowest-$u$ vs highest-$u$ Chicago, $t_{26} = 1.54$, $p > 0.10$; and lowest-$u$ vs highest-$u$ US, $t_{26} = 1.13$, $p > 0.10$.

If we focus on the product categories where the sample size is smaller for the lowest unemployment period, then among the 25 such cases, the difference in asymmetry threshold is in the right direction in 21 cases; it is the same in 1 case (in the cereals category, the lowest-$u$ period vs. the highest-$u$ US), and it is in the opposite direction in 3 cases (in the cereals category: the lowest-$u$ period vs. the highest-$u$ Chicago, and in the toothbrush category: the lowest-$u$ period vs. the highest-$u$ Chicago, and the lowest-$u$ period vs. the highest-$u$ US). The average asymmetry threshold is significantly bigger in the lowest-$u$ period ($\bar{A} = 11.52$, SD = 7.37) than in the highest-$u$ periods ($\bar{A} = 3.24$, SD = 4.32; $t_{24} = 3.93$, $p < 0.001$).[8]

We repeated the same comparison for each of the three highest-unemployment sample periods separately. Among the 5 cases where the sample size is smaller for the lowest unemployment period than the NBER recession period, the average asymmetry is larger in the lowest-$u$ period ($\bar{A} = 16$, SD = 5.92 vs $\bar{A} = 0.40$, SD = 0.55, $t_4 = 5.96$, $p = 0.002$). Among the 10 cases where the sample size is smaller for the lowest-$u$ period than the highest-$u$ Chicago, the average asymmetry is bigger in the lowest-$u$ period, but the difference is not statistically significant ($\bar{A} = 10.40$, SD = 10.4 vs $\bar{A} = 5.10$, SD = 5.95, $t_9 = 1.34$, $p = 0.21$). Among the 10 cases where the sample size is smaller for the lowest-$u$ period than the

---

[8] $\bar{A} = 11.52$ is the average of the asymmetry thresholds for the 25 low-unemployment cases. $\bar{A} = 3.24$ is the average of the three asymmetry thresholds for the high unemployment periods: NBER recession, Highest-$u$ Chicago, and Highest-$u$ US.

highest-*u* US, the average asymmetry is significantly bigger in the lowest-*u* period ($\bar{A}$ =10.40, SD = 8.07 vs $\bar{A}$ = 2.80, SD = 2.78, $t_9$ = 2.64, $p < 0.03$). We conclude that the differences in asymmetry thresholds are unlikely to be driven by differences in sample sizes.

Could our findings be an artifact of the negative short-run inflation-unemployment relationship? In that case, the finding of smaller asymmetry thresholds during the highest unemployment would imply that there was deflation during that specific period. We therefore look at the inflation rates for the three 8-month periods, using the PPI, CPI-US, and CPI-Chicago (Table A1 in the Appendix). If we compare the 8 months with the highest unemployment to the 8 months with the lowest unemployment, the average inflation rate is higher for the former (for all three indices). The results are the same if we compare the 8 months with the lowest unemployment with the NBER recession period. The unemployment rate is higher in the recession, but the inflation rate is also higher (for all three indices). We conclude that our findings are unrelated to the inflation-unemployment relationship.

Another explanation might be that in high-unemployment periods, there are more price cuts if retailers try to boost sales in economic downturns. However, price cuts should not affect asymmetry thresholds systematically because in Dominick's data, they are temporary and therefore reversed (Rotemberg 2005, Chen et al 2008, Midrigan 2011). For robustness, we explored this by identifying sales events using a V-shaped sales filter of Syed (2015) and Fox and Syed (2016) and excluding them from the analysis.[9]

The figures in Table A2 in the appendix indicate that the main results are unaffected. The asymmetry threshold for the lowest unemployment period, $\bar{A}$ = 10.85¢, is larger than for the NBER recession period $\bar{A}$ = 0.48¢, for the Highest-*u* Chicago period $\bar{A}$ = 3.78¢, and for the Highest-*u* US period $\bar{A}$ = 2.81¢. Across the 27 categories, in 56 out of 75 comparisons, i.e., in 74.67% of cases, we find a stronger asymmetry for the lowest unemployment period, in 5 cases (6.7% of the cases) equal asymmetry, and in 14 cases (18.67% of the cases) weaker asymmetry for the lowest unemployment period than for the other periods.

We also considered the effects of clearance sales on our results by identifying instances where products are withdrawn following a price cut of 10% or more and excluding them from the analysis. As the figures in Table A3 in the appendix show, the asymmetry thresholds we obtain here are no different from what we report in Table 1.

We rerun the analysis by simultaneously excluding both the V-shaped sales events and clearance sales. The results reported in Table A4 in the appendix are similar to what we report in Table A2 in the appendix.

---

[9] The literature offers about a dozen different sales filters (Sandler et al. 2024). The filter of Syed (2015) and Fox and Syed (2016) is calibrated for Dominick's price data, making it particularly useful for us. The filter is used by Snir and Levy (2021) and Snir et al. (2022). Note that we used a sales filter rather than Dominick's sale indicator dummy because the latter was not set consistently (Peltzman 2000).

We conclude that excluding V-shaped sales and clearance sales does not alter our key results: In most cases, the asymmetry in small price changes is stronger in booms than in busts, as hypothesized.

## 7. Conclusion, caveats, and future work

Chen et al. (2008) explain asymmetric price adjustment in small price changes by consumer inattention to small price changes, giving the sellers the incentive to make more frequent small price increases than decreases. Recent studies find a correlation between unemployment and shoppers' attention to prices: the higher the unemployment, the more attentive shoppers are to prices. These two sets of findings lead to a testable prediction: we should see a variation in the extent of asymmetry in small price adjustments over the business cycle. During booms (busts), because the unemployment rate is low (high) and shoppers are less (more) attentive, we should see greater (lesser) asymmetry in small price adjustments.

We use large scanner data for 1989–1997, which includes an 8-month period of NBER recession, to compare the asymmetry in small price changes between the recession (high $u$) and expansion (low $u$) periods. Consistent with our hypothesis, we find a greater asymmetry in small price changes during low-unemployment periods, compared with high-unemployment periods, implying that firm price-setting behavior varies over the business cycle.

We shall note, however, that we study a single (although quite large) chain over a period that contains a single cycle of low and high unemployment. It will be useful, therefore, if future work further explores these questions using different datasets over other periods of recessions and expansions, to provide additional evidence of how asymmetry in small price adjustments varies over the business cycle.

**Acknowledgments**

We are grateful to the anonymous reviewer for constructive comments and suggestions, which helped us in improving the manuscript, and to the editor, Eric Young, for guidance. We thank various conference and seminar participants for their helpful discussions. The views expressed herein are those of the authors and not necessarily those of the Federal Reserve Bank of Minneapolis or the Federal Reserve System. All errors are ours.

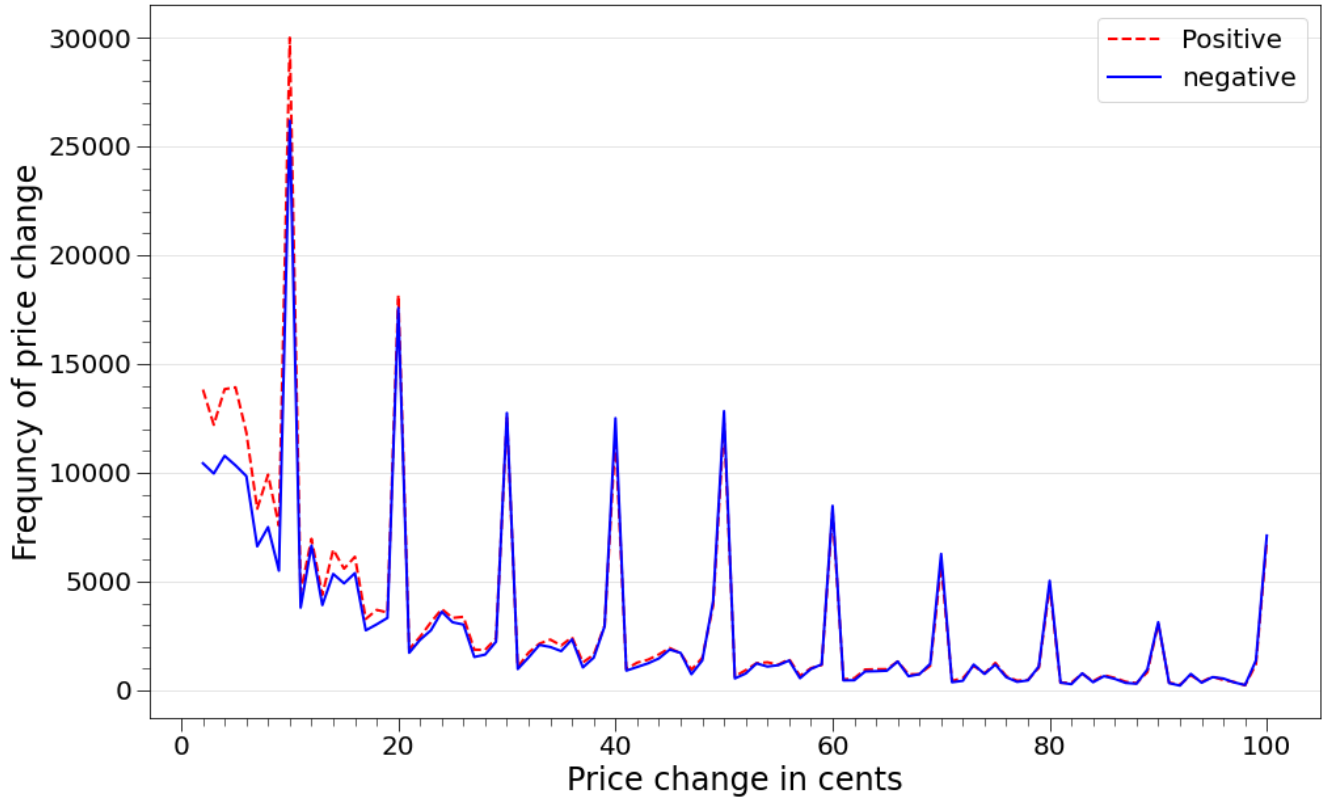

**Fig. 1.** Average frequency of positive and negative price changes, all 29 categories

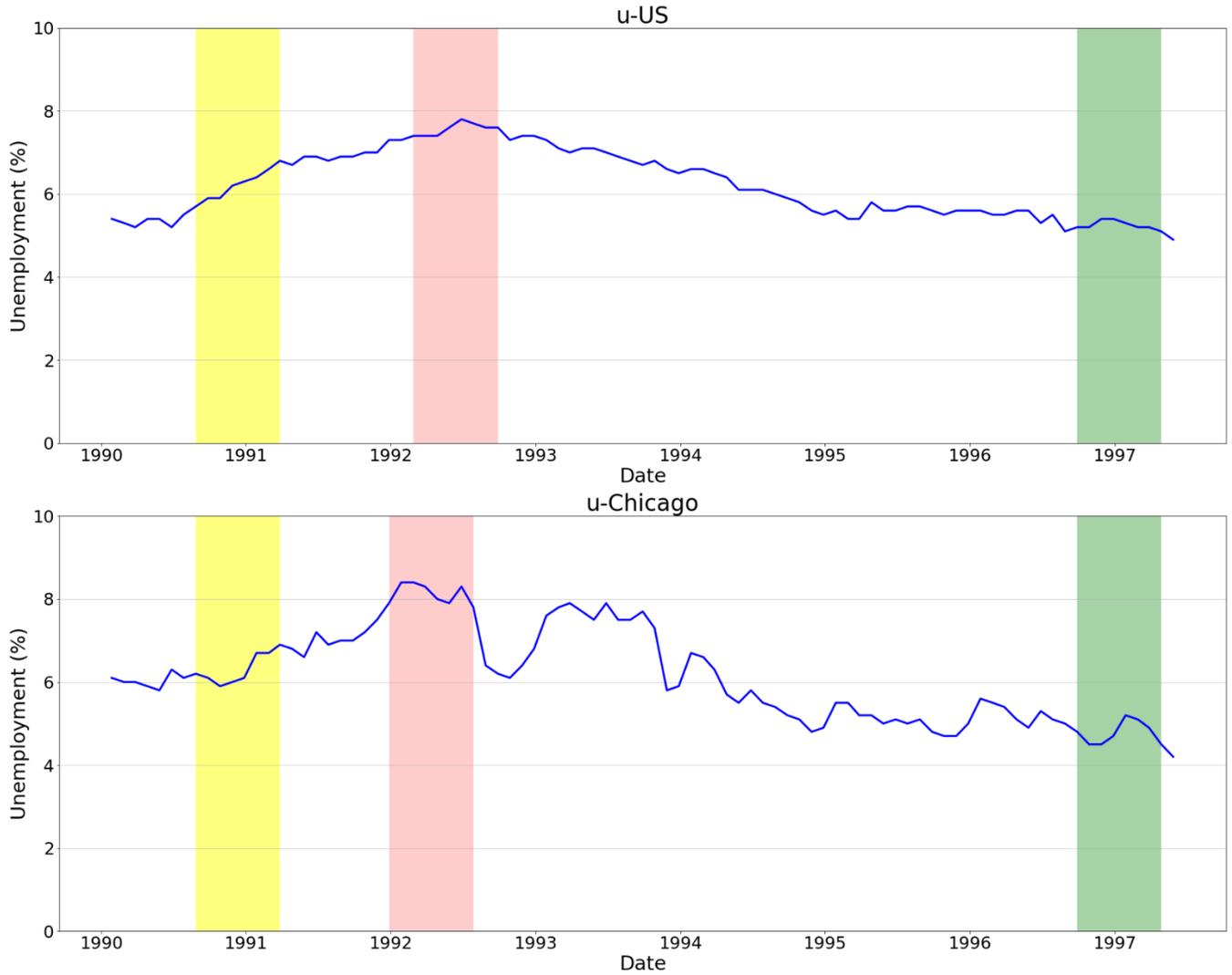

**Fig. 2.** Monthly unemployment rate in the US and Chicago, 1989–1997

Notes:
1. The left-hand side shaded area marks the NBER recession period, August 1990–March 1991.
2. The middle-shaded area marks the highest unemployment rate periods, February 1992–September 1992 in the US, and December 1991–July 1992 in Chicago.
3. The right-hand side shaded area marks the lowest unemployment periods, September 1996 –April 1997, in both the US and Chicago.

**Table 1**
Variation in the asymmetry threshold in cents over the business cycle

| Product Categories | Asymmetry threshold ($\bar{A}$) in cents | | | | Sample size | | | |
|---|---|---|---|---|---|---|---|---|
| | Lowest $u$ | NBER Recession | Highest-$u$ Chicago | Highest-$u$ US | Lowest $u$ | NBER Recession | Highest-$u$ Chicago | Highest-$u$ US |
| Analgesics | 16 | 0 | 8 | 8 | 290,098 | 243,554 | 275,751 | 271,589 |
| Bath Soap | 0 | -- | 0 | 4 | 66,850 | -- | 29,693 | 40,445 |
| Bathroom Tissues | 4 | 3 | 1 | 1 | 119,928 | 81,772 | 95,866 | 97,704 |
| Bottled Juices | 12 | 2 | 6 | 6 | 396,630 | 296,436 | 398,069 | 400,885 |
| Canned Soup | 17 | 0 | 1 | 1 | 270,074 | 480,363 | 510,137 | 513,003 |
| Canned Tuna | 25 | 1 | 2 | 2 | 169,238 | 204,450 | 225,749 | 229,596 |
| Cereals | 0 | 0 | 20 | 0 | 444,826 | 435,170 | 465,991 | 469,343 |
| Cheeses | 29 | 0 | 1 | 1 | 640,023 | 545,066 | 590,552 | 594,712 |
| Cookies | 9 | 1 | 8 | 6 | 629,269 | 658,658 | 720,327 | 724,924 |
| Crackers | 11 | 0 | 1 | 1 | 267,978 | 184,937 | 198,575 | 194,353 |
| Dish Detergent | 15 | 0 | 2 | 2 | 208,650 | 192,674 | 191,233 | 191,155 |
| Fabric Softeners | 1 | 0 | 4 | 3 | 195,268 | 180,544 | 190,898 | 193,299 |
| Front-end-candies | 16 | 0 | 1 | 1 | 339,746 | 391,849 | 409,466 | 414,510 |
| Frozen Dinners | 9 | -- | 7 | 3 | 219,267 | -- | 52,357 | 104,752 |
| Frozen Entrees | 19 | 0 | 11 | 8 | 666,595 | 595,097 | 626,024 | 627,971 |
| Frozen Juices | 10 | 0 | 2 | 1 | 200,042 | 190,792 | 209,811 | 211,856 |
| Grooming Products | 18 | -- | 10 | 10 | 686,463 | -- | 292,428 | 408,529 |
| Laundry Detergents | 13 | 0 | 2 | 2 | 239,687 | 256,294 | 301,483 | 304,595 |
| Oatmeal | 4 | -- | 0 | 18 | 116,311 | -- | 112,143 | 107,397 |
| Paper Towels | 2 | 0 | 1 | 1 | 81,136 | 73,354 | 84,240 | 83,448 |
| Refrigerated Juices | 15 | 0 | 10 | 1 | 207,171 | 149,588 | 177,756 | 176,872 |
| Shampoos | 17 | -- | 3 | 6 | 816,157 | -- | 493,778 | 683,457 |
| Snack Crackers | 3 | 1 | 2 | 2 | 309,361 | 297,408 | 301,817 | 304,149 |
| Soaps | 9 | -- | 1 | 1 | 226,417 | -- | 183,734 | 214,697 |
| Soft Drinks | 3 | 3 | 0 | 0 | 1,262,488 | 658,506 | 774,846 | 791,416 |
| Toothbrushes | 0 | 0 | 8 | 8 | 168,467 | 162,515 | 187,868 | 192,626 |
| Toothpastes | 1 | 2 | 0 | 0 | 294,654 | 238,442 | 251,899 | 252,323 |
| Average Threshold ($\bar{A}$) | 10.30 | 0.62 | 4.15 | 3.59 | | | | |
| Median | 10 | 0 | 2 | 2 | | | | |

Notes:
1. Lowest $u$ denotes the lowest unemployment rate period for both the City of Chicago and the US.
2. NBER Recession denotes the NBER recession period.

3. Highest-*u* Chicago denotes the highest Chicago unemployment rate period.
4. Highest-*u* US denotes the highest U.S. unemployment rate period.
5. The empty cells are the cases of missing observations.

# Online Supplementary Web Appendix

**(Not for Publication)**

# Asymmetric Price Adjustment over the Business Cycle

Daniel Levy*
Department of Economics
Bar-Ilan University, Emory University, ICEA, ISET at TSU, and RCEA

Haipeng (Allan) Chen
Tippie College of Business
University of Iowa

Sourav Ray
Lang School of Business and Economics
University of Guelph

Elliot Charette
Department of Applied Economics
University of Minnesota, and
Federal Reserve Bank of Minneapolis

Xiao Ling
School of Business
Central Connecticut State University

Weihong Zhao
Robert H. Smith School of Business
University of Maryland

Mark Bergen
Carlson School of Management
University of Minnesota

Avichai Snir
Department of Economics
Bar-Ilan University

Last Revision: June 11, 2025

* Corresponding author: Daniel Levy, Daniel.Levy@biu.ac.il

**Table A1**

Three measures of inflation (*PPI*, *CPI*, and *CPI*-Chicago) and two measures of unemployment (*u*-US and *u*-Chicago), September 1989–May 1997

| Year | Month | PPI | %ΔPPI | CPI | %ΔCPI | CPI-Chicago | %ΔCPI-Chicago | u-US | u-Chicago |
|---|---|---|---|---|---|---|---|---|---|
| 1989 | September | 113.6 | - | 125.0 | - | 127.1 | - | 5.3 | - |
| 1989 | October | 114.9 | 1.14 | 125.6 | 0.5 | 126.8 | –0.2 | 5.3 | - |
| 1989 | November | 114.9 | 0.00 | 125.9 | 0.2 | 126.7 | –0.1 | 5.4 | - |
| 1989 | December | 115.4 | 0.44 | 126.1 | 0.2 | 126.5 | –0.2 | 5.4 | - |
| 1990 | January | 117.6 | 1.91 | 127.4 | 1.0 | 128.1 | 1.3 | 5.4 | 6.1 |
| 1990 | February | 117.4 | –0.17 | 128.0 | 0.5 | 129.2 | 0.9 | 5.3 | 6.0 |
| 1990 | March | 117.2 | –0.17 | 128.7 | 0.5 | 129.5 | 0.2 | 5.2 | 6.0 |
| 1990 | April | 117.2 | 0.00 | 128.9 | 0.2 | 130.4 | 0.7 | 5.4 | 5.9 |
| 1990 | May | 117.7 | 0.43 | 129.2 | 0.2 | 130.4 | 0.0 | 5.4 | 5.8 |
| 1990 | June | 117.8 | 0.08 | 129.9 | 0.5 | 131.7 | 1.0 | 5.2 | 6.3 |
| 1990 | July | 118.2 | 0.34 | 130.4 | 0.4 | 132.0 | 0.2 | 5.5 | 6.1 |
| 1990 | August | 119.3 | 0.93 | 131.6 | 0.9 | 133.2 | 0.9 | 5.7 | 6.2 |
| 1990 | September | 120.4 | 0.92 | 132.7 | 0.8 | 133.8 | 0.5 | 5.9 | 6.1 |
| 1990 | October | 122.3 | 1.58 | 133.5 | 0.6 | 133.3 | –0.4 | 5.9 | 5.9 |
| 1990 | November | 122.9 | 0.49 | 133.8 | 0.2 | 134.2 | 0.7 | 6.2 | 6.0 |
| 1990 | December | 122.0 | –0.73 | 133.8 | 0.0 | 134.6 | 0.3 | 6.3 | 6.1 |
| 1991 | January | 122.3 | 0.25 | 134.6 | 0.6 | 135.1 | 0.4 | 6.4 | 6.7 |
| 1991 | February | 121.4 | –0.74 | 134.8 | 0.1 | 135.5 | 0.3 | 6.6 | 6.7 |
| 1991 | March | 120.9 | –0.41 | 135.0 | 0.1 | 136.2 | 0.5 | 6.8 | 6.9 |
| 1991 | April | 121.1 | 0.17 | 135.2 | 0.1 | 136.1 | –0.1 | 6.7 | 6.8 |
| 1991 | May | 121.8 | 0.58 | 135.6 | 0.3 | 136.8 | 0.5 | 6.9 | 6.6 |
| 1991 | June | 121.9 | 0.08 | 136.0 | 0.3 | 137.3 | 0.4 | 6.9 | 7.2 |
| 1991 | July | 121.6 | –0.25 | 136.2 | 0.1 | 137.3 | 0.0 | 6.8 | 6.9 |
| 1991 | August | 121.7 | 0.08 | 136.6 | 0.3 | 137.6 | 0.2 | 6.9 | 7.0 |
| 1991 | September | 121.4 | –0.25 | 137.2 | 0.4 | 138.3 | 0.5 | 6.9 | 7.0 |
| 1991 | October | 122.2 | 0.66 | 137.4 | 0.1 | 138.0 | –0.2 | 7.0 | 7.2 |
| 1991 | November | 122.3 | 0.08 | 137.8 | 0.3 | 138.0 | 0.0 | 7.0 | 7.5 |
| 1991 | December | 121.9 | –0.33 | 137.9 | 0.1 | 138.3 | 0.2 | 7.3 | 7.9 |
| 1992 | January | 121.8 | –0.08 | 138.1 | 0.1 | 138.9 | 0.4 | 7.3 | 8.4 |
| 1992 | February | 122.1 | 0.25 | 138.6 | 0.4 | 139.2 | 0.2 | 7.4 | 8.4 |
| 1992 | March | 122.2 | 0.08 | 139.3 | 0.5 | 139.7 | 0.4 | 7.4 | 8.3 |
| 1992 | April | 122.4 | 0.16 | 139.5 | 0.1 | 139.8 | 0.1 | 7.4 | 8.0 |
| 1992 | May | 123.2 | 0.65 | 139.7 | 0.1 | 140.5 | 0.5 | 7.6 | 7.9 |
| 1992 | June | 123.9 | 0.57 | 140.2 | 0.4 | 141.2 | 0.5 | 7.8 | 8.3 |
| 1992 | July | 123.7 | –0.16 | 140.5 | 0.2 | 141.4 | 0.1 | 7.7 | 7.8 |
| 1992 | August | 123.6 | –0.08 | 140.9 | 0.3 | 141.9 | 0.4 | 7.6 | 6.4 |
| 1992 | September | 123.3 | –0.24 | 141.3 | 0.3 | 142.7 | 0.6 | 7.6 | 6.2 |
| 1992 | October | 124.4 | 0.89 | 141.8 | 0.4 | 142.1 | –0.4 | 7.3 | 6.1 |
| 1992 | November | 124.0 | –0.32 | 142.0 | 0.1 | 142.4 | 0.2 | 7.4 | 6.4 |
| 1992 | December | 123.8 | –0.16 | 141.9 | –0.1 | 142.9 | 0.4 | 7.4 | 6.8 |
| 1993 | January | 124.2 | 0.32 | 142.6 | 0.5 | 143.2 | 0.2 | 7.3 | 7.6 |
| 1993 | February | 124.5 | 0.24 | 143.1 | 0.4 | 143.6 | 0.3 | 7.1 | 7.8 |
| 1993 | March | 124.7 | 0.16 | 143.6 | 0.3 | 144.1 | 0.3 | 7.0 | 7.9 |
| 1993 | April | 125.5 | 0.64 | 144.0 | 0.3 | 144.7 | 0.4 | 7.1 | 7.7 |
| 1993 | May | 125.8 | 0.24 | 144.2 | 0.1 | 145.7 | 0.7 | 7.1 | 7.5 |
| 1993 | June | 125.5 | –0.24 | 144.4 | 0.1 | 145.6 | –0.1 | 7.0 | 7.9 |
| 1993 | July | 125.3 | –0.16 | 144.4 | 0.0 | 145.5 | –0.1 | 6.9 | 7.5 |
| 1993 | August | 124.2 | –0.88 | 144.8 | 0.3 | 146.1 | 0.4 | 6.8 | 7.5 |

| Year | Month | | | | | | | | |
|---|---|---|---|---|---|---|---|---|---|
| 1993 | September | 123.8 | −0.32 | 145.1 | 0.2 | 146.7 | 0.4 | 6.7 | 7.7 |
| 1993 | October | 124.6 | 0.65 | 145.7 | 0.4 | 147.2 | 0.3 | 6.8 | 7.3 |
| 1993 | November | 124.5 | −0.08 | 145.8 | 0.1 | 146.4 | −0.5 | 6.6 | 5.8 |
| 1993 | December | 124.1 | −0.32 | 145.8 | 0.0 | 146.1 | −0.2 | 6.5 | 5.9 |
| 1994 | January | 124.5 | 0.32 | 146.2 | 0.3 | 146.5 | 0.3 | 6.6 | 6.7 |
| 1994 | February | 124.8 | 0.24 | 146.7 | 0.3 | 146.8 | 0.2 | 6.6 | 6.6 |
| 1994 | March | 124.9 | 0.08 | 147.2 | 0.3 | 147.6 | 0.5 | 6.5 | 6.3 |
| 1994 | April | 125.0 | 0.08 | 147.4 | 0.1 | 147.9 | 0.2 | 6.4 | 5.7 |
| 1994 | May | 125.3 | 0.24 | 147.5 | 0.1 | 147.6 | −0.2 | 6.1 | 5.5 |
| 1994 | June | 125.6 | 0.24 | 148.0 | 0.3 | 148.1 | 0.3 | 6.1 | 5.8 |
| 1994 | July | 126.0 | 0.32 | 148.4 | 0.3 | 148.3 | 0.1 | 6.1 | 5.5 |
| 1994 | August | 126.5 | 0.40 | 149.0 | 0.4 | 149.8 | 1.0 | 6.0 | 5.4 |
| 1994 | September | 125.6 | −0.71 | 149.4 | 0.3 | 150.2 | 0.3 | 5.9 | 5.2 |
| 1994 | October | 125.8 | 0.16 | 149.5 | 0.1 | 149.4 | −0.5 | 5.8 | 5.1 |
| 1994 | November | 126.1 | 0.24 | 149.7 | 0.1 | 150.4 | 0.7 | 5.6 | 4.8 |
| 1994 | December | 126.2 | 0.08 | 149.7 | 0.0 | 150.5 | 0.1 | 5.5 | 4.9 |
| 1995 | January | 126.6 | 0.32 | 150.3 | 0.4 | 151.8 | 0.9 | 5.6 | 5.5 |
| 1995 | February | 126.9 | 0.24 | 150.9 | 0.4 | 152.3 | 0.3 | 5.4 | 5.5 |
| 1995 | March | 127.1 | 0.16 | 151.4 | 0.3 | 152.6 | 0.2 | 5.4 | 5.2 |
| 1995 | April | 127.6 | 0.39 | 151.9 | 0.3 | 153.1 | 0.3 | 5.8 | 5.2 |
| 1995 | May | 128.1 | 0.39 | 152.2 | 0.2 | 153.0 | −0.1 | 5.6 | 5.0 |
| 1995 | June | 128.2 | 0.08 | 152.5 | 0.2 | 153.5 | 0.3 | 5.6 | 5.1 |
| 1995 | July | 128.2 | 0.00 | 152.5 | 0.0 | 153.6 | 0.1 | 5.7 | 5.0 |
| 1995 | August | 128.1 | −0.08 | 152.9 | 0.3 | 153.8 | 0.1 | 5.7 | 5.1 |
| 1995 | September | 127.9 | −0.16 | 153.2 | 0.2 | 154.0 | 0.1 | 5.6 | 4.8 |
| 1995 | October | 128.7 | 0.63 | 153.7 | 0.3 | 154.3 | 0.2 | 5.5 | 4.7 |
| 1995 | November | 128.7 | 0.00 | 153.6 | −0.1 | 154.0 | −0.2 | 5.6 | 4.7 |
| 1995 | December | 129.1 | 0.31 | 153.5 | −0.1 | 153.8 | −0.1 | 5.6 | 5.0 |
| 1996 | January | 129.4 | 0.23 | 154.4 | 0.6 | 154.6 | 0.5 | 5.6 | 5.6 |
| 1996 | February | 129.4 | 0.00 | 154.9 | 0.3 | 155.2 | 0.4 | 5.5 | 5.5 |
| 1996 | March | 130.1 | 0.54 | 155.7 | 0.5 | 156.3 | 0.7 | 5.5 | 5.4 |
| 1996 | April | 130.6 | 0.38 | 156.3 | 0.4 | 156.4 | 0.1 | 5.6 | 5.1 |
| 1996 | May | 131.1 | 0.38 | 156.6 | 0.2 | 156.9 | 0.3 | 5.6 | 4.9 |
| 1996 | June | 131.7 | 0.46 | 156.7 | 0.1 | 157.6 | 0.4 | 5.3 | 5.3 |
| 1996 | July | 131.5 | −0.15 | 157.0 | 0.2 | 157.7 | 0.1 | 5.5 | 5.1 |
| 1996 | August | 131.9 | 0.30 | 157.3 | 0.2 | 158.1 | 0.3 | 5.1 | 5.0 |
| 1996 | September | 131.8 | −0.08 | 157.8 | 0.3 | 158.3 | 0.1 | **5.2** | **4.8** |
| 1996 | October | 132.7 | 0.68 | 158.3 | 0.3 | 158.8 | 0.3 | **5.2** | **4.5** |
| 1996 | November | 132.6 | −0.08 | 158.6 | 0.2 | 159.4 | 0.4 | **5.4** | **4.5** |
| 1996 | December | 132.7 | 0.08 | 158.6 | 0.0 | 159.7 | 0.2 | **5.4** | **4.7** |
| 1997 | January | 132.6 | −0.08 | 159.1 | 0.3 | 160.4 | 0.4 | **5.3** | **5.2** |
| 1997 | February | 132.2 | −0.30 | 159.6 | 0.3 | 161.1 | 0.4 | **5.2** | **5.1** |
| 1997 | March | 132.1 | −0.08 | 160.0 | 0.3 | 161.0 | −0.1 | **5.2** | **4.9** |
| 1997 | April | 131.6 | −0.38 | 160.2 | 0.1 | 160.9 | −0.1 | **5.1** | **4.5** |
| 1997 | May | 131.6 | 0.00 | 160.1 | −0.1 | 161.1 | 0.1 | 4.9 | 4.2 |

Notes

**Yellow: August 1990–March 1991** - NBER Recession Months - 8 months

**Navy: September 1996–April 1997** - Chicago - Lowest unemployment rate 8-month period, $\bar{u} = 4.8\%$

**Green: September 1996–April 1997** - US - Lowest unemployment rate 8-month period, $\bar{u} = 5.2\%$

**Blue: February 1992–September 1992** - US - Highest unemployment rate 8-month period, $\bar{u} = 7.6\%$

**Purple: December 1991–July 1992** - Chicago - Highest unemployment rate 8-month period, $\bar{u} = 8.1\%$

**Table A2**

Variation in the asymmetry thresholds in cents over the business cycle, after excluding V-shaped sales events

| | Asymmetry threshold ($\bar{A}$) | | | | Sample size | | | |
|---|---|---|---|---|---|---|---|---|
| Product Categories | Lowest $u$ | NBER Recession | Highest-$u$ Chicago | Highest-$u$ US | Lowest $u$ | NBER Recession | Highest-$u$ Chicago | Highest-$u$ US |
| Analgesics | 0 | 0 | 8 | 8 | 247,027 | 194,952 | 256,879 | 254,113 |
| Bath Soap | 0 | -- | 0 | 4 | 62,201 | -- | 25,841 | 34,529 |
| Bathroom Tissues | 3 | 0 | 0 | 0 | 100,972 | 60,383 | 60,969 | 63,685 |
| Bottled Juices | 32 | 2 | 0 | 0 | 335,060 | 230,481 | 293,839 | 291,778 |
| Canned Soup | 16 | 0 | 0 | 0 | 220,292 | 410,157 | 443,668 | 447,123 |
| Canned Tuna | 10 | 1 | 0 | 0 | 149,901 | 159,343 | 176,287 | 184,184 |
| Cereals | 0 | 0 | 21 | 0 | 399,987 | 397,768 | 427,337 | 398,884 |
| Cheeses | 32 | 0 | 0 | 0 | 510,315 | 398,904 | 428,277 | 433,358 |
| Cookies | 20 | 0 | 0 | 0 | 480,465 | 527,293 | 547,184 | 536,090 |
| Crackers | 23 | 0 | 0 | 0 | 198,120 | 123,084 | 141,456 | 140,516 |
| Dish Detergent | 12 | 0 | 0 | 0 | 180,224 | 153,348 | 158,044 | 157,446 |
| Fabric Softeners | 1 | 0 | 0 | 0 | 170,769 | 144,974 | 157,918 | 154,171 |
| Front-end-candies | 5 | 0 | 1 | 0 | 312,058 | 368,840 | 364,181 | 375,411 |
| Frozen Dinners | 1 | -- | 18 | 5 | 156,394 | -- | 35,949 | 68,925 |
| Frozen Entrees | 10 | 0 | 11 | 9 | 479,724 | 457,468 | 453,459 | 448,185 |
| Frozen Juices | 14 | 0 | 1 | 0 | 155,118 | 141,441 | 157,220 | 158,265 |
| Grooming Products | 18 | -- | 0 | 10 | 535,251 | -- | 234,010 | 330,053 |
| Laundry Detergents | 10 | 0 | 1 | 1 | 203,769 | 214,815 | 240,892 | 239,516 |
| Oatmeal | 4 | -- | 0 | 14 | 98,187 | -- | 102,935 | 101,272 |
| Paper Towels | 1 | 0 | 0 | 0 | 68,403 | 50,007 | 58,732 | 58,362 |
| Refrigerated Juices | 10 | 0 | 11 | 0 | 155,013 | 88,334 | 104,080 | 102,328 |
| Shampoos | 10 | -- | 3 | 8 | 622,122 | -- | 396,415 | 543,336 |
| Snack Crackers | 2 | 1 | 1 | 1 | 228,789 | 214,795 | 201,254 | 221,418 |
| Soaps | 11 | -- | 0 | 0 | 197,827 | -- | 146,843 | 173,222 |
| Soft Drinks | 48 | 3 | 13 | 3 | 740,748 | 382,807 | 428,307 | 444,675 |
| Toothbrushes | 0 | 0 | 8 | 8 | 115,268 | 132,646 | 167,911 | 166,280 |
| Toothpastes | 0 | 3 | 5 | 5 | 225,329 | 193,596 | 217,500 | 211,898 |
| Average Threshold ($\bar{A}$) | 10.85 | 0.48 | 3.78 | 2.81 | | | | |
| Median | 10 | 0 | 0 | 0 | | | | |

Notes:
1. Lowest $u$ denotes the lowest unemployment rate period, for both the City of Chicago and the U.S.
2. NBER Recession denotes the NBER recession period.
3. Highest-$u$ Chicago denotes the highest Chicago unemployment rate period.
4. Highest-$u$ US denotes the highest U.S. unemployment rate period.
5. The blank cells are the cases of missing observations

**Table A3**
Variation in the asymmetry thresholds in cents over the business cycle, after excluding clearance sales events

|  | Asymmetry threshold ($\bar{A}$) | | | | Sample size | | | |
|---|---|---|---|---|---|---|---|---|
| Product Categories | Lowest $u$ | NBER Recession | Highest-$u$ Chicago | Highest-$u$ US | Lowest $u$ | NBER Recession | Highest-$u$ Chicago | Highest-$u$ US |
| Analgesics | 16 | 0 | 8 | 8 | 290,042 | 243,467 | 275,733 | 271,554 |
| Bath Soap | 0 | -- | 0 | 4 | 66,836 | -- | 29,681 | 40,433 |
| Bathroom Tissues | 4 | 3 | 1 | 1 | 119,903 | 81,770 | 95,849 | 97,684 |
| Bottled Juices | 12 | 2 | 6 | 6 | 396,211 | 296,320 | 398,011 | 400,841 |
| Canned Soup | 17 | 0 | 1 | 1 | 269,942 | 480,339 | 510,137 | 513,003 |
| Canned Tuna | 25 | 1 | 2 | 2 | 169,220 | 204,371 | 225,702 | 229,552 |
| Cereals | 0 | 0 | 20 | 0 | 444,768 | 435,156 | 465,938 | 469,266 |
| Cheeses | 29 | 0 | 1 | 1 | 639,059 | 545,034 | 590,486 | 594,571 |
| Cookies | 9 | 1 | 8 | 6 | 629,066 | 658,632 | 720,165 | 724,731 |
| Crackers | 11 | 0 | 1 | 1 | 267,800 | 184,936 | 198,548 | 194,290 |
| Dish Detergent | 15 | 0 | 2 | 2 | 208,618 | 192,469 | 191,221 | 191,147 |
| Fabric Softeners | 1 | 0 | 4 | 3 | 195,080 | 180,434 | 190,736 | 193,136 |
| Front-end-candies | 16 | 0 | 1 | 1 | 339,727 | 391,759 | 409,453 | 414,472 |
| Frozen Dinners | 9 | -- | 7 | 3 | 219,161 | -- | 52,326 | 104,702 |
| Frozen Entrees | 19 | 0 | 11 | 8 | 665,977 | 594,446 | 625,524 | 627,418 |
| Frozen Juices | 10 | 0 | 2 | 1 | 200,032 | 190,737 | 209,737 | 211,780 |
| Grooming Products | 18 | -- | 10 | 10 | 685,873 | -- | 292,341 | 408,311 |
| Laundry Detergents | 13 | 0 | 2 | 2 | 239,502 | 255,708 | 301,281 | 304,449 |
| Oatmeal | 4 | -- | 0 | 18 | 116,309 | -- | 112,142 | 107,396 |
| Paper Towels | 2 | 0 | 1 | 1 | 81,125 | 73,332 | 84,184 | 83,394 |
| Refrigerated Juices | 15 | 0 | 10 | 1 | 207,081 | 149,455 | 177,736 | 176,825 |
| Shampoos | 17 | -- | 3 | 6 | 815,200 | -- | 493,357 | 682,806 |
| Snack Crackers | 3 | 1 | 2 | 2 | 309,279 | 297,382 | 301,704 | 304,060 |
| Soaps | 9 | -- | 1 | 1 | 226,341 | -- | 183,608 | 214,570 |
| Soft Drinks | 3 | 3 | 0 | 0 | 1,260,976 | 658,340 | 774,370 | 790,879 |
| Toothbrushes | 0 | 0 | 8 | 8 | 168,319 | 162,481 | 187,857 | 192,607 |
| Toothpastes | 1 | 2 | 0 | 0 | 294,477 | 238,394 | 251,796 | 252,000 |
| Average Threshold ($\bar{A}$) | 10.30 | 0.62 | 4.15 | 3.59 | | | | |
| Median | 10 | 0 | 2 | 2 | | | | |

Notes:
1. Lowest $u$ denotes the lowest unemployment rate period, for both the City of Chicago and the U.S.
2. NBER Recession denotes the NBER recession period.
3. Highest-$u$ Chicago denotes the highest Chicago unemployment rate period.
4. Highest-$u$ US denotes the highest U.S. unemployment rate period.
5. The blank cells are the cases of missing observations

**Table A4**
Variation in the asymmetry thresholds in cents over the business cycle, after simultaneously excluding V-shaped sales and clearance sales events

| Product Categories | Asymmetry threshold ($\bar{A}$) | | | | Sample size | | | |
|---|---|---|---|---|---|---|---|---|
| | Lowest $u$ | NBER Recession | Highest-$u$ Chicago | Highest-$u$ US | Lowest $u$ | NBER Recession | Highest-$u$ Chicago | Highest-$u$ US |
| Analgesics | 0 | 0 | 8 | 8 | 247,027 | 194,952 | 256,877 | 254,111 |
| Bath Soap | 0 | -- | 0 | 4 | 62,201 | -- | 25,841 | 34,529 |
| Bathroom Tissues | 3 | 0 | 0 | 0 | 100,972 | 60,383 | 60,969 | 63,685 |
| Bottled Juices | 32 | 2 | 0 | 0 | 335,060 | 230,481 | 293,839 | 291,778 |
| Canned Soup | 16 | 0 | 0 | 0 | 220,292 | 410,154 | 443,668 | 447,123 |
| Canned Tuna | 10 | 1 | 0 | 0 | 149,901 | 159,336 | 176,285 | 184,184 |
| Cereals | 0 | 0 | 21 | 0 | 399,987 | 397,768 | 427,337 | 398,882 |
| Cheeses | 32 | 0 | 0 | 0 | 510,315 | 398,904 | 428,271 | 433,352 |
| Cookies | 20 | 0 | 0 | 0 | 480,465 | 527,289 | 547,179 | 536,086 |
| Crackers | 23 | 0 | 0 | 0 | 198,120 | 123,084 | 141,455 | 140,508 |
| Dish Detergent | 12 | 0 | 0 | 0 | 180,224 | 153,348 | 158,043 | 157,446 |
| Fabric Softeners | 1 | 0 | 0 | 0 | 170,769 | 144,972 | 157,905 | 154,158 |
| Front-end-candies | 5 | 0 | 1 | 0 | 312,058 | 368,832 | 364,181 | 375,411 |
| Frozen Dinners | 1 | -- | 18 | 5 | 156,394 | -- | 35,946 | 68,921 |
| Frozen Entrees | 10 | 0 | 11 | 9 | 479,724 | 457,455 | 453,442 | 448,152 |
| Frozen Juices | 14 | 0 | 1 | 0 | 155,118 | 141,441 | 157,220 | 158,265 |
| Grooming Products | 18 | -- | 0 | 10 | 535,251 | -- | 234,005 | 330,044 |
| Laundry Detergents | 10 | 0 | 1 | 1 | 203,769 | 214,787 | 240,890 | 239,514 |
| Oatmeal | 4 | -- | 0 | 14 | 98,187 | -- | 102,935 | 101,272 |
| Paper Towels | 1 | 0 | 0 | 0 | 68,403 | 50,003 | 58,727 | 58,359 |
| Refrigerated Juices | 10 | 0 | 11 | 0 | 155,013 | 88,334 | 104,080 | 102,328 |
| Shampoos | 10 | -- | 3 | 8 | 622,122 | -- | 396,398 | 543,317 |
| Snack Crackers | 2 | 1 | 1 | 1 | 228,789 | 214,795 | 201,247 | 221,411 |
| Soaps | 11 | -- | 0 | 0 | 197,827 | -- | 146,835 | 173,214 |
| Soft Drinks | 48 | 3 | 13 | 3 | 740,748 | 382,802 | 428,281 | 444,646 |
| Toothbrushes | 0 | 0 | 8 | 8 | 115,268 | 132,645 | 167,909 | 166,278 |
| Toothpastes | 0 | 3 | 5 | 5 | 225,329 | 193,596 | 217,499 | 211,897 |
| Average Threshold ($\bar{A}$) | 10.85 | 0.48 | 3.78 | 2.81 | | | | |
| Median | 10 | 0 | 0 | 0 | | | | |

Notes:
1. Lowest $u$ denotes the lowest unemployment rate period, for both the City of Chicago and the U.S.
2. NBER Recession denotes the NBER recession period.
3. Highest-$u$ Chicago denotes the highest Chicago unemployment rate period.
4. Highest-$u$ US denotes the highest U.S. unemployment rate period.
5. The blank cells are the cases of missing observations